\documentclass[12]{article}
\usepackage{amssymb}
\usepackage{graphics} 
\usepackage[lmargin=1.7in,rmargin=1.5in,tmargin=1.4in,bmargin=1.2in]{geometry}
\begin{document}
\mbox{}\\[1mm]
\begin{center}
{\LARGE \bf A finite-dimensional quantum model\\[2mm] for the stock market}\\[5mm]
{Liviu-Adrian Cotfas}\\[5mm]
{\it Faculty of Economic Cybernetics, Statistics and Informatics,\\ Academy of Economic Studies,  Bucharest, Romania\\ E-mail:   \texttt{lcotfas@gmail.com}}
\end{center}
\bigskip
{\bf Abstract}. We present a finite-dimensional version of the quantum model for the stock market  proposed in  [C. Zhang and L. Huang, A quantum model for the stock market,  Physica A  389 (2010) 5769]. Our approach is an attempt to make this model consistent with the discrete nature of the stock price and is based on the mathematical formalism used in the case of  the quantum systems with finite-dimensional Hilbert space. The rate of return is a discrete variable corresponding to the coordinate in the case of quantum systems, and the operator of the conjugate variable describing the trend of the stock return is defined in terms of the finite Fourier transform. The stock return in equilibrium is described by a finite Gaussian function, and the time evolution of the stock price, directly related to the rate of return, is obtained by numerically solving  a Schr\" odinger type equation.
\bigskip
\section{Introduction}
\label{introd}

The methods of quantum mechanics are among the tools used in order to get a deeper insight in 
  the complexity of the financial markets [1-9]. The time evolution of the price and the trend of the stock in the financial markets depends on many factors including the political environment, market information, economic policies of the government, psychology of traders, etc.
Prices at which traders are willing to buy or sell a financial asset are not more determined by the development of industry, trade, services, situation at the market of natural resources and so on \cite{Choustova}. The information exchange and market psychology play important role in price dynamics.

The mathematical modeling of price dynamics of the financial market is a very complex problem. We could never take into account all economic and non-economic conditions that have influences to the market. Therefore, we usually consider some very simplified and idealized models, a kind of toy models which mimic certain features of a real stock market. For a better understanding of  the complexity of the financial markets one has to use 
several complementary models. 

The variables describing the market dynamics are discrete variables, not continuous ones.
For example, the stock price at a given moment of time is an integer multiple of a certain minimal quantity, a sort of quantum of cash (usually, $1/100$ or $1/1000$ of the currency unit) \cite{Bagarello2006}. The time evolution of the price is thus described by a step function. The quantum models facilitate the description of phenomena not fully explained by the classical theory of prices. The lack of simultaneously observability of certain variables and the interference of attempted values noted by traders and economists represent such quantum effects. There is the impossibility of observing prices and their instantaneous forward time derivative even if we consider price as a continuous variable \cite{Segal}. The rate of return being directly related to the price, there is also the impossibility of observing the rate of return and  its instantaneous forward time derivative. The trade of a stock can be regarded as the basic process that measures its momentary price. The stock price can only be determined at the time of sale when it is between traders. We can never simultaneously know both the ownership of a stock and its price \cite{Zhang, Pedram}. 
The stock price $\wp $ is a variable taking numerical values but the ownership seems to be a very different kind of variable.

Our aim is to present a quantum model describing the time evolution of the rate of return.
We shall take into consideration only the case of stock markets with  a price limit rule: the rate of return 
\begin{equation}
\mathcal{R}=\frac{\wp -\wp _0}{\wp _0}=\frac{\wp }{\wp _0}-1
\end{equation}
in a trading day can not be more than $\pm q\%$ comparing with the previous day's closing price $\wp _0$, that is, we have 
\[
-\frac{q}{100}\leq \mathcal{R}\leq \frac{q}{100}.
\]
For example \cite{Zhang}, in most Chinese stock markets $q=10$.
In their activity, the traders do not take into consideration an arbitrary small rate of return. They approximate the rate of return by integer multiples of a minimal one.
We shall consider $\mathcal{R}$ as a discrete variable with the only possible values
\[
\begin{array}{l}
-\frac{q}{100}, \ \ -\frac{q\!-\!1}{100},\ \ ...\ \ , -\frac{1}{100}, \ \ 0,\ \ \frac{1}{100}, \ \ ...\ \ , \frac{q\!-\!1}{100},\ \ \frac{q}{100}
\end{array}
\]
and by following the analogy with quantum mechanics, we describe the rate of return at a fixed moment of time by a wave function
\[
\begin{array}{l}
\Phi :\left\{ \frac{-q}{100},\, \frac{-q+1}{100},\, ...\, ,\, \frac{q-1}{100},\, \frac{q}{100}\right\}\longrightarrow \mathbb{C}
\end{array}
\]
such that $\left| \Phi \left(\frac{n}{100}\right)\right|^2$ is the probability to have a return rate equal to $\frac{n}{100}$.
%
%
\section{Infinite-dimensional quantum system}
\label{iquantsys}

In the case of a quantum particle moving along a straight line, the possible positions form the set $\mathbb{R}\!=\!(-\infty ,\infty )$, and each state of the system is described by a function $\psi \!:\!\mathbb{R}\!\longrightarrow \!\mathbb{C}$ satisfying the relation (and called square integrable)
\[
\int_{-\infty }^\infty |\psi (x)|^2\, dx<\infty .
\]
The space of square integrable functions considered with the scalar product
\begin{equation}
\langle \psi _1,\psi _2\rangle =\int_{-\infty }^\infty \overline{\psi _1(x)}\, \psi _2(x)\, dx
\end{equation}
is an infinite dimensional Hilbert space,  denoted by $L^2(\mathbb{R})$. To each function $\psi \!:\!\mathbb{R}\!\longrightarrow \!\mathbb{C}$ we associate its Fourier transform $\mathcal{F}[\psi ]\!:\!\mathbb{R}\!\longrightarrow \!\mathbb{C}$ defined as
\begin{equation}
\mathcal{F}[\psi ](p)=\frac{1}{\sqrt{2\pi \hbar}}\int_{-\infty }^\infty {\rm e}^{-{\rm i}px/\hbar }\, \psi (x)\, dx 
\end{equation}
where $\hbar $ is Planck's constant $h$ divided by $2\pi $.
The transformation $\psi \mapsto \mathcal{F}[\psi ]$ is a unitary transformation
\[
\langle \mathcal{F}[\psi _1], \mathcal{F}[\psi _2]\rangle =\langle \psi _1,\psi _2\rangle \qquad {\rm for\ any}\quad \psi _1, \psi _2 \in L^2(\mathbb{R})
\]
and its inverse is the transformation 
\begin{equation}
\mathcal{F}^{-1}[\psi ](x)=\frac{1}{\sqrt{2\pi \hbar}}\int_{-\infty }^\infty {\rm e}^{{\rm i}px/\hbar }\, \psi (p)\, dp. 
\end{equation}
The normalized function corresponding to a function $\psi \neq 0$, namely,
\[
\begin{array}{l}
\Psi \!:\!\mathbb{R}\!\longrightarrow \!\mathbb{C}, \qquad \Psi (x)=\frac{1}{\sqrt{\langle \psi ,\psi \rangle }}\, \psi (x)
\end{array}
\]
satisfies the relation
\[
\int_{-\infty }^\infty |\Psi (x)|^2\, dx =1.
\]
In the state described by the normalized function $\Psi $, the number
\[
\int_a^b |\Psi (x)|^2\, dx 
\]
represents the probability to find the particle in the interval $[a,b]$, and
 \[
\int_a^b |\mathcal{F}[\Psi ](p)|^2\, dp 
\]
represents the probability to have a momentum in the interval $[a,b]$.\\
Each observable $A$ is described by a linear operator $\hat A $ satisfying the relation
\begin{equation}
\langle \hat A\psi _1,\psi _2\rangle =\langle \psi _1,\hat A \psi _2\rangle 
\end{equation}
called a Hermitian operator. The position is described by the operator \cite{Messiah}
\begin{equation}
 \psi \mapsto \hat x \, \psi\qquad  {\rm where}\qquad  (\hat x \, \psi )(x)=x\, \psi (x)
\end{equation}
and the momentum by the operator 
\begin{equation}
\psi \mapsto \hat p \, \psi \qquad  {\rm where}\qquad  \hat p \, \psi =-{\rm i}\hbar  \frac{d\psi }{dx}.
\end{equation}
The position and momentum operators  satisfy the relation
\begin{equation}\label{momop}
\hat p=\mathcal{F}^{-1}\hat x\mathcal{F}.
\end{equation}
In a state described by the  normalized wavefunction $\Psi $, the numbers 
\[
\langle \hat x\rangle =\langle \Psi ,\hat x\, \Psi \rangle \qquad {\rm and} \qquad 
\langle \hat p\rangle =\langle \Psi ,\hat p\, \Psi \rangle 
\]
represent the mean value of coordinate and momentum, respectively.
The time evolution of the state of the system satisfies the Schr\" odiger equation
\begin{equation}
\begin{array}{l}
{\rm i}\, \frac{\partial }{\partial t}\, \Psi (x,t)=\left( \frac{1}{2\mu }\hat p^2+\mathcal{V}(\hat x,t)\right)\Psi (x,t)
\end{array}
\end{equation}
where $\mu $ is the mass of the particle and the operator $\mathcal{V}(\hat x,t)$ is obtained from the potential energy $\mathcal{V}(x,t)$ by using the substitution $x\mapsto \hat x$.
%
%
\section{Finite-dimensional quantum systems}
\label{quantsys}

 We obtain a very simplified version of the infinite-dimensional quantum  system presented in the previous section by assuming that we can distinguish only a finite number $d$ of positions for our particle. 
It is convenient to consider that $d\!=\!2s\!+\!1$ is an odd number and to assume that 
\begin{equation}
\begin{array}{l}
\mathcal{S}_d=\left\{ -s\sqrt{\frac{2\pi }{d}}, \, (-s\!+\!1)\sqrt{\frac{2\pi }{d}},\, ...\, , \, (s\!-\!1)\sqrt{\frac{2\pi }{d}},\, s\sqrt{\frac{2\pi }{d}}\right\}
\end{array}
\end{equation}
is the set of all the positions we can distinguish. From the relations
\[
\begin{array}{l}
\lim\limits_{d\rightarrow \infty }\sqrt{\frac{2\pi }{d}}\!=\!0,\qquad  \lim\limits_{d\rightarrow \infty }(-s)\sqrt{\frac{2\pi }{d}}\!=\!-\infty ,\qquad \lim\limits_{d\rightarrow \infty }s\sqrt{\frac{2\pi }{d}}\!=\!\infty 
\end{array}
\]
it follows that, in a certain sense, we have $\lim\limits_{d\rightarrow \infty }\mathcal{S}_d=(-\infty ,\infty )$.

Each state of our system is described by a function $\psi \!:\!\mathcal{S}_d\!\longrightarrow \! \mathbb{C}$. The space $\mathcal{H}$ of all these functions considered with the scalar product
\begin{equation}
\langle \psi _1,\psi _2\rangle =\sum_{x\in \mathcal{S}_d} \overline{\psi _1(x)}\, \psi _2(x)
\end{equation}
is a $d$-dimensional Hilbert space isomorphic to $\mathbb{C}^d$. To each function $\psi \!:\!\mathcal{S}_d\!\rightarrow \! \mathbb{C}$ we associate its Fourier transform $F[\psi ]\!:\!\mathcal{S}_d\!\longrightarrow \! \mathbb{C}$ defined as \cite{Vourdas,Mehta}
\[
F[\psi ](p)=\frac{1}{\sqrt{d}}\sum_{x\in \mathcal{S}_d}{\rm e}^{-{\rm i}px}\, \psi (x) 
\]
that is, 
\begin{equation}
\begin{array}{l}
F[\psi ]\left(k\sqrt{\frac{2\pi }{d}}\right)=\frac{1}{\sqrt{d}}\sum\limits_{n=-s}^s{\rm e}^{-\frac{2\pi {\rm i}}{d}kn}\, \psi \left(n\sqrt{\frac{2\pi }{d}}\right) .
\end{array}
\end{equation}
The transformation $\psi \mapsto F[\psi ]$ is a unitary transformation
\[
\langle F[\psi _1], F[\psi _2]\rangle =\langle \psi _1,\psi _2\rangle \qquad {\rm for\ any}\quad \psi _1, \psi _2 \in \mathcal{H}
\]
and its inverse is the transformation 
\begin{equation}
\begin{array}{l}
F^{-1}[\psi ]\left(n\sqrt{\frac{2\pi }{d}}\right)=\frac{1}{\sqrt{d}}\sum\limits_{k=-s}^s{\rm e}^{\frac{2\pi {\rm i}}{d}kn}\, \psi \left(k\sqrt{\frac{2\pi }{d}}\right) .
\end{array}
\end{equation}
The normalized function corresponding to a function $\psi \neq 0$, namely,
\begin{equation}
\begin{array}{l}
\Psi \!:\!\mathcal{S}_d\!\longrightarrow \!\mathbb{C}, \qquad \Psi (x)=\frac{1}{\sqrt{\langle \psi ,\psi \rangle }}\, \psi (x)
\end{array}
\end{equation}
satisfies the relation
\begin{equation}
\begin{array}{l}
\sum\limits_{n=-s}^s\left| \Psi \left(n\sqrt{\frac{2\pi }{d}}\right)\right|^2=1. 
\end{array}
\end{equation}
In the state described by the normalized function $\Psi $, the number
\[
\begin{array}{l}
\left| \Psi \left(n\sqrt{\frac{2\pi }{d}}\right)\right|^2
\end{array}
\]
represents the probability to find the particle at point $n\sqrt{\frac{2\pi }{d}}$, and the number
\[
\begin{array}{l}
\left| F[\Psi ]\left(k\sqrt{\frac{2\pi }{d}}\right)\right|^2
\end{array}
\]
the probability to have a momentum equal to $k\sqrt{\frac{2\pi }{d}}$.
Each observable $A$ is described by a linear operator $\hat A $ satisfying the relation
\[
\langle \hat A\psi _1,\psi _2\rangle =\langle \psi _1,\hat A \psi _2\rangle\qquad {\rm for\ any}\quad \psi _1, \psi _2 \in \mathcal{H}
\]
called a Hermitian operator. The position is described by the operator
\[
\hat Q:\mathcal{H}\longrightarrow \mathcal{H}:\psi\mapsto  \hat Q\psi 
\]
where
\[
(\hat Q\psi ) (x)=x\, \psi (x)
\]
that is,
\begin{equation} \label{defq}
\begin{array}{l}
 (\hat Q\psi ) \left(n\sqrt{\frac{2\pi }{d}}\right)=n\sqrt{\frac{2\pi }{d}}\, \, \psi \!\left(n\sqrt{\frac{2\pi }{d}}\right) .
\end{array}
\end{equation}
The function $\psi $ is a function of discrete variable and the expression $\frac{d\psi }{dx}$ is meaningless. Therefore, we define the operator corresponding to the momentum  by following the analogy with the relation (\ref{momop}) as
\begin{equation}
\hat P:\mathcal{H}\longrightarrow \mathcal{H}, \qquad \hat P=F^{-1}\hat QF.
\end{equation}
In a state described by the  normalized wavefunction $\Psi $, the numbers 
\[
\begin{array}{l}
\langle \hat Q\rangle =\langle \Psi ,\hat Q\, \Psi \rangle =\sum\limits_{n=-s}^sn\sqrt{\frac{2\pi }{d}}\, \left| \Psi \left(n\sqrt{\frac{2\pi }{d}}\right)\right|^2
\end{array}
\]
and
\[
\begin{array}{l}
\langle \hat P\rangle =\langle \Psi ,\hat P\, \Psi \rangle =\langle \Psi ,F^{-1}\hat QF\, \Psi \rangle=\langle F[\Psi ],\hat QF[ \Psi ]\rangle \\[3mm]
\mbox{}\qquad \qquad \qquad \ \,   =\sum\limits_{k=-s}^sk\sqrt{\frac{2\pi }{d}}\, \left| F[\Psi ]\left(k\sqrt{\frac{2\pi }{d}}\right)\right|^2
\end{array}
\]
represent the mean value of coordinate and momentum, respectively.
The time evolution of the state of the system satisfies a Schr\" odiger type equation
\begin{equation}
{\rm i}\, \frac{\partial }{\partial t}\, \Psi =\hat H\, \Psi .
\end{equation}
where the operator
\begin{equation}
\hat H=\frac{1}{2\mu }\hat P^2+\mathcal{V}(\hat Q,t)
\end{equation}
is the corresponding Hamiltonian. If $\hat H$ does not depend on time than the solution of the Schr\" odinger equation satisfies the relation
\begin{equation}  
\begin{array}{l}
\Psi \left(n\sqrt{\frac{2\pi }{d}},t\right)={\rm e}^{-it\hat H}\Psi \left(n\sqrt{\frac{2\pi }{d}},0\right).
\end{array}
\end{equation}
The finite-dimensional  quantum system presented in this section converges for $d\rightarrow \infty $ to the system presented in the previous section, in a certain sense.

\section{A quantum model for the stock market}
\label{model}

We consider a stock market with a price limit rule \cite{Zhang}, for which the rate of return $\mathcal{R}$ in a trading day can not be more than $\pm q\%$. By assuming that
\begin{equation}
\begin{array}{l}
\mathcal{L}_d=\left\{ \frac{-q}{100},\, \frac{-q+1}{100},\, ...\, ,\, \frac{q-1}{100},\, \frac{q}{100}\right\}
\end{array}
\end{equation}
is the set of all the possible values of the rate of return we use a function $\varphi :\mathcal{L}_d\longrightarrow \mathbb{C}$ in order to describe  the rate of return.
The space $\mathcal{K}$ of all the functions $\varphi :\mathcal{L}_d\longrightarrow \mathbb{C}
$ considered with the scalar product
\begin{equation}
\begin{array}{l}
\langle \varphi _1|\varphi _2\rangle =\sum\limits_{n=-q}^q\overline{\varphi _1\left(\frac{n}{100}\right)}\, \, \varphi _2\left(\frac{n}{100}\right)
\end{array}
\end{equation}
is a Hilbert space of dimension $d\!=\!2q\!+\!1$, isomorphic to $\mathbb{C}^d$.
To each function $\varphi \!:\!\mathcal{L}_d\!\rightarrow \! \mathbb{C}$ we associate its Fourier transform $F[\varphi ]\!:\!\mathcal{L}_d\!\longrightarrow \! \mathbb{C}$ defined as 
\begin{equation}
\begin{array}{l}
F[\varphi ]\left(\frac{k}{100}\right)=\frac{1}{\sqrt{d}}\sum\limits_{n=-q}^q{\rm e}^{-\frac{2\pi {\rm i}}{d}kn}\, \varphi \left(\frac{n}{100}\right) .
\end{array}
\end{equation}
The transformation $\varphi \mapsto F[\varphi ]$ is a unitary transformation
\[
\langle F[\varphi _1], F[\varphi _2]\rangle =\langle \varphi _1,\varphi _2\rangle \qquad {\rm for\ any}\quad \varphi _1, \varphi _2 \in \mathcal{K}
\]
and its inverse is the transformation 
\begin{equation}
\begin{array}{l}
F^{-1}[\varphi ]\left(\frac{n}{100}\right)=\frac{1}{\sqrt{d}}\sum\limits_{k=-q}^q{\rm e}^{\frac{2\pi {\rm i}}{d}kn}\, \varphi \left(\frac{k}{100}\right) .
\end{array}
\end{equation}
The normalized function corresponding to a function $\varphi \neq0$, namely,
\[
\begin{array}{l}
\Phi :\mathcal{L}_d\longrightarrow \mathbb{C},\qquad \Phi \left(\frac{n}{100}\right)=\frac{1}{\sqrt{\langle \varphi, \varphi\rangle }}\, \varphi  \left(\frac{n}{100}\right).
\end{array}
\]
satisfies the relation
\begin{equation}
\begin{array}{l}
\sum\limits_{n=-q}^q\left|\Phi \left(\frac{n}{100}\right)\right|^2=1.
\end{array}
\end{equation}
By following the analogy with quantum mechanics, we assume that in the case of a rate of return described by the normalized function $\Phi $, the number
\[
\begin{array}{l}
\left|\Phi \left(\frac{n}{100}\right)\right|^2
\end{array}
\]
represents the probability to have a return rate equal to $\frac{n}{100}$.

In our model the financial variables are described by Hermitian operators. The rate of return corresponds to the position, and by following the analogy with quantum mechanics we define the corresponding operator as
\begin{equation} \label{defq}
\begin{array}{l}
\hat R:\mathcal{K}\longrightarrow \mathcal{K},\qquad 
 (\hat R\varphi ) \left(\frac{n}{100}\right)=\frac{n}{100}\, \, \varphi \!\left(\frac{n}{100}\right) .
\end{array}
\end{equation}
Mean value of the rate of return $\mathcal{R}$ in the case of a normalized function $\Phi $ is
\[
\begin{array}{l}
\langle \hat R\rangle =\langle \Phi ,\hat R\Phi \rangle =\sum\limits_{n=-q}^q\frac{n}{100}\,  \left|\Phi \left(\frac{n}{100}\right)\right|^2.
\end{array}
\]

The rate of return has been defined by using the stock price as 
\[
\mathcal{R}=\frac{\wp -\wp _0}{\wp _0}=\frac{\wp }{\wp _0}-1.
\]
This relation, written in the form
\[
\wp=\wp_0+\wp_0\, \mathcal{R}
\]
allows us to define the price operator as
\begin{equation}
\hat \wp =\wp _0\mathbb{I}+\wp _0\, \hat R
\end{equation}
where $\mathbb{I}:\mathcal{K}\longrightarrow \mathcal{K}$, \, $\mathbb{I}\varphi =\varphi $ is the identity operator.

In our discrete version, we cannot define the operator $\hat T$ corresponding to the trend of return rate by using the method from \cite{Zhang} as a differential operator. We define it directly by using the finite Fourier transform as
\begin{equation}
\hat T :\mathcal{K}\longrightarrow \mathcal{K},\qquad \hat T=F^{-1}\hat RF.
\end{equation}

The set of functions $\{ \delta _n \}_{n=-q}^q$, where 
\begin{equation}
\begin{array}{l}
\delta _n: \mathcal{L}_d\longrightarrow \mathbb{C},\qquad 
\delta _n\left(\frac{m}{100}\right)=\left\{ 
\begin{array}{lll}
1 & {\rm if}&  n=m\\
0 & {\rm if}& n\neq m
\end{array} \right.
\end{array}
\end{equation}
is an orthonormal basis in $\mathcal{K}$, called the canonical basis. The function $\delta _n$ is an eigenfunction of the operator  $\hat R$ corresponding to the eigenvalue $\frac{n}{100}$
\begin{equation}
\begin{array}{l}
\hat R\,  \delta _n =\frac{n}{100}\, \delta _n 
\end{array}
\end{equation}
and any function $\varphi :\mathcal{L}_d\longrightarrow \mathbb{C}$ can be written as a linear combination of $\delta _n$
\begin{equation}
\begin{array}{l}
\varphi =\sum\limits_{n=-q}^q\varphi \!\left(\frac{n}{100}\right)\, \delta _n.
\end{array}
\end{equation}
The set of functions $\{ \Phi _n \}_{n=-q}^q$ defined as $\Phi _n=F^{-1}[\delta _n]$, that is, as
\begin{equation}
\begin{array}{l}
\Phi _n: \mathcal{L}_d\longrightarrow \mathbb{C},\qquad 
\Phi _n\left(\frac{k}{100}\right)=\frac{1}{\sqrt{d}}{\rm e}^{\frac{2\pi {\rm i}}{d}kn}
\end{array}
\end{equation}
is a second orthonormal basis in $\mathcal{K}$. Each  $\Phi _n$ is an eigenfunction of $\hat T$
\begin{equation}\label{Teigen}
\begin{array}{l}
\hat T\, \Phi _n=F^{-1}\hat RFF^{-1}[\delta _n]=F^{-1}\hat R\delta _n=\frac{n}{100}\, F^{-1}[\delta _n ]=\frac{n}{100}\,\Phi _n
\end{array}
\end{equation}
and any function $\varphi :\mathcal{L}_d\longrightarrow \mathbb{C}$ can be written as a linear combination of $\Phi _n$
\[
\begin{array}{l}
\varphi =\sum\limits_{n=-q}^q\langle \Phi _n,\varphi \rangle \, \Phi _n.
\end{array}
\]
Since 
\[
\begin{array}{l}
\langle \Phi _n,\varphi \rangle =\langle F^{-1}[\delta _n],\varphi \rangle =\langle \delta _n,F[\varphi ]\rangle =F[\varphi ]\left(\frac{n}{100}\right)
\end{array}
\]
each function $\varphi :\mathcal{L}_d\longrightarrow \mathbb{C}$ admits the representation
\begin{equation}\label{Trepr}
\begin{array}{l}
\varphi =\sum\limits_{n=-q}^qF[\varphi ]\left(\frac{n}{100}\right) \, \Phi _n.
\end{array}
\end{equation}
In the case of a normalized function $\Phi $ we have
\[
\begin{array}{l}
\sum\limits_{k=-q}^q\left|F[\Phi ]\left(\frac{k}{100}\right)\right|^2=\sum\limits_{k=-q}^q|\langle \Phi _k,\Phi \rangle |^2=1
\end{array}
\]
and the number
\[
\begin{array}{l}
\left|F[\Phi ]\left(\frac{k}{100}\right)\right|^2
\end{array}
\]
is the probability to have a trend equal to $\frac{k}{100}$.

By following the analogy with the quantum systems, we assume that the function $\Phi $ describing the rate of return satisfies a Schr\" odinger type equation
\begin{equation}
{\rm i}\, \frac{\partial }{\partial t}\, \Phi (x,t)=\hat H\, \Phi (x,t).
\end{equation}
where the operator
\begin{equation}
\hat H=\frac{1}{2\mu }\hat P^2+\mathcal{V}(\hat Q,t)
\end{equation}
is the corresponding Hamiltonian. By denoting
\[
\phi _n=\langle \delta _n,\Phi \rangle \qquad {\rm and}\qquad h_{n,m}=\langle \delta _n,\hat H\delta _m\rangle
\]
the Schr\" odinger equation can be written in the matrix form
\[
{\rm i}\frac{d}{dt}\left(
\begin{array}{l}
\phi _{-s}(t)\\
\phi _{-s+1}(t)\\
\vdots \\
\phi _{s}(t)
\end{array} \right)\!=\!
\left(
\begin{array}{llll}
h_{-s,-s} & h_{-s,-s+1} & \dots & h_{-s,s}\\
h_{-s+1,-s} & h_{-s+1,-s+1} & \dots & h_{-s+1,s}\\
\vdots & \vdots &\ddots & \vdots \\
h_{s,-s} & h_{s,-s+1} & \dots & h_{s,s}
\end{array}\right)
\left(
\begin{array}{l}
\phi _{-s}(t)\\
\phi _{-s+1}(t)\\
\vdots \\
\phi _{s}(t)
\end{array} \right) 
\]
that is, it is a system of $d$ first order differential equations with $d$ unknown functions $\phi _{-s}$, $\phi _{-s+1}$, ... , $\phi _{s}$.

\begin{center}
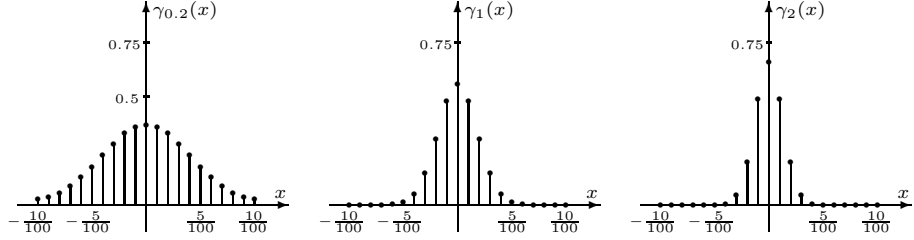
\begin{figure}[t]
\setlength{\unitlength}{1.8mm}
\begin{picture}(70,20)(-5,0)
\put(0,2){\vector(1,0){20}}
\put(23,2){\vector(1,0){20}}
\put(46,2){\vector(1,0){20}}
\put(9.5,0){\vector(0,1){17}}
\put(32.5,0){\vector(0,1){17}}
\put(55.5,0){\vector(0,1){17}}

\put(    1.50000,   2.45684){\circle*{0.3}}
\put(    2.30000,   2.60584){\circle*{0.3}}
\put(    3.10000,   2.91252){\circle*{0.3}}
\put(    3.90000,   3.38704){\circle*{0.3}}
\put(    4.70000,   4.02870){\circle*{0.3}}
\put(    5.50000,   4.81235){\circle*{0.3}}
\put(    6.30000,   5.67880){\circle*{0.3}}
\put(    7.10000,   6.53499){\circle*{0.3}}
\put(    7.90000,   7.26650){\circle*{0.3}}
\put(    8.70000,   7.76099){\circle*{0.3}}
\put(    9.50000,   7.93595){\circle*{0.3}}
\put(   10.30000,   7.76099){\circle*{0.3}}
\put(   11.10000,   7.26650){\circle*{0.3}}
\put(   11.90000,   6.53499){\circle*{0.3}}
\put(   12.70000,   5.67880){\circle*{0.3}}
\put(   13.50000,   4.81235){\circle*{0.3}}
\put(   14.30000,   4.02870){\circle*{0.3}}
\put(   15.10000,   3.38704){\circle*{0.3}}
\put(   15.90000,   2.91252){\circle*{0.3}}
\put(   16.70000,   2.60584){\circle*{0.3}}
\put(   17.50000,   2.45684){\circle*{0.3}}
\put(    1.50000,2){\line(0,1){    .45684}}
\put(    2.30000,2){\line(0,1){    .60584}}
\put(    3.10000,2){\line(0,1){    .91252}}
\put(    3.90000,2){\line(0,1){   1.38704}}
\put(    4.70000,2){\line(0,1){   2.02870}}
\put(    5.50000,2){\line(0,1){   2.81235}}
\put(    6.30000,2){\line(0,1){   3.67880}}
\put(    7.10000,2){\line(0,1){   4.53499}}
\put(    7.90000,2){\line(0,1){   5.26650}}
\put(    8.70000,2){\line(0,1){   5.76099}}
\put(    9.50000,2){\line(0,1){   5.93595}}
\put(   10.30000,2){\line(0,1){   5.76099}}
\put(   11.10000,2){\line(0,1){   5.26650}}
\put(   11.90000,2){\line(0,1){   4.53499}}
\put(   12.70000,2){\line(0,1){   3.67880}}
\put(   13.50000,2){\line(0,1){   2.81235}}
\put(   14.30000,2){\line(0,1){   2.02870}}
\put(   15.10000,2){\line(0,1){   1.38704}}
\put(   15.90000,2){\line(0,1){    .91252}}
\put(   16.70000,2){\line(0,1){    .60584}}
\put(   17.50000,2){\line(0,1){    .45684}}
\put(   24.50000,   2.00000){\circle*{0.3}}
\put(   25.30000,   2.00000){\circle*{0.3}}
\put(   26.10000,   2.00062){\circle*{0.3}}
\put(   26.90000,   2.00582){\circle*{0.3}}
\put(   27.70000,   2.04073){\circle*{0.3}}
\put(   28.50000,   2.21114){\circle*{0.3}}
\put(   29.30000,   2.81152){\circle*{0.3}}
\put(   30.10000,   4.31254){\circle*{0.3}}
\put(   30.90000,   6.88587){\circle*{0.3}}
\put(   31.70000,   9.65336){\circle*{0.3}}
\put(   32.50000,  10.88838){\circle*{0.3}}
\put(   33.30000,   9.65336){\circle*{0.3}}
\put(   34.10000,   6.88587){\circle*{0.3}}
\put(   34.90000,   4.31254){\circle*{0.3}}
\put(   35.70000,   2.81152){\circle*{0.3}}
\put(   36.50000,   2.21114){\circle*{0.3}}
\put(   37.30000,   2.04073){\circle*{0.3}}
\put(   38.10000,   2.00582){\circle*{0.3}}
\put(   38.90000,   2.00062){\circle*{0.3}}
\put(   39.70000,   2.00000){\circle*{0.3}}
\put(   40.50000,   2.00000){\circle*{0.3}}
\put(   24.50000,2){\line(0,1){    .00000}}
\put(   25.30000,2){\line(0,1){    .00000}}
\put(   26.10000,2){\line(0,1){    .00062}}
\put(   26.90000,2){\line(0,1){    .00582}}
\put(   27.70000,2){\line(0,1){    .04073}}
\put(   28.50000,2){\line(0,1){    .21114}}
\put(   29.30000,2){\line(0,1){    .81152}}
\put(   30.10000,2){\line(0,1){   2.31254}}
\put(   30.90000,2){\line(0,1){   4.88587}}
\put(   31.70000,2){\line(0,1){   7.65336}}
\put(   32.50000,2){\line(0,1){   8.88838}}
\put(   33.30000,2){\line(0,1){   7.65336}}
\put(   34.10000,2){\line(0,1){   4.88587}}
\put(   34.90000,2){\line(0,1){   2.31254}}
\put(   35.70000,2){\line(0,1){    .81152}}
\put(   36.50000,2){\line(0,1){    .21114}}
\put(   37.30000,2){\line(0,1){    .04073}}
\put(   38.10000,2){\line(0,1){    .00582}}
\put(   38.90000,2){\line(0,1){    .00062}}
\put(   39.70000,2){\line(0,1){    .00000}}
\put(   40.50000,2){\line(0,1){    .00000}}
\put(   47.50000,   2.00000){\circle*{0.3}}
\put(   48.30000,   2.00000){\circle*{0.3}}
\put(   49.10000,   2.00000){\circle*{0.3}}
\put(   49.90000,   2.00000){\circle*{0.3}}
\put(   50.70000,   2.00022){\circle*{0.3}}
\put(   51.50000,   2.00596){\circle*{0.3}}
\put(   52.30000,   2.08811){\circle*{0.3}}
\put(   53.10000,   2.71551){\circle*{0.3}}
\put(   53.90000,   5.19387){\circle*{0.3}}
\put(   54.70000,   9.83682){\circle*{0.3}}
\put(   55.50000,  12.57013){\circle*{0.3}}
\put(   56.30000,   9.83682){\circle*{0.3}}
\put(   57.10000,   5.19387){\circle*{0.3}}
\put(   57.90000,   2.71551){\circle*{0.3}}
\put(   58.70000,   2.08811){\circle*{0.3}}
\put(   59.50000,   2.00596){\circle*{0.3}}
\put(   60.30000,   2.00022){\circle*{0.3}}
\put(   61.10000,   2.00000){\circle*{0.3}}
\put(   61.90000,   2.00000){\circle*{0.3}}
\put(   62.70000,   2.00000){\circle*{0.3}}
\put(   63.50000,   2.00000){\circle*{0.3}}
\put(   47.50000,2){\line(0,1){    .00000}}
\put(   48.30000,2){\line(0,1){    .00000}}
\put(   49.10000,2){\line(0,1){    .00000}}
\put(   49.90000,2){\line(0,1){    .00000}}
\put(   50.70000,2){\line(0,1){    .00022}}
\put(   51.50000,2){\line(0,1){    .00596}}
\put(   52.30000,2){\line(0,1){    .08811}}
\put(   53.10000,2){\line(0,1){    .71551}}
\put(   53.90000,2){\line(0,1){   3.19387}}
\put(   54.70000,2){\line(0,1){   7.83682}}
\put(   55.50000,2){\line(0,1){  10.57013}}
\put(   56.30000,2){\line(0,1){   7.83682}}
\put(   57.10000,2){\line(0,1){   3.19387}}
\put(   57.90000,2){\line(0,1){    .71551}}
\put(   58.70000,2){\line(0,1){    .08811}}
\put(   59.50000,2){\line(0,1){    .00596}}
\put(   60.30000,2){\line(0,1){    .00022}}
\put(   61.10000,2){\line(0,1){    .00000}}
\put(   61.90000,2){\line(0,1){    .00000}}
\put(   62.70000,2){\line(0,1){    .00000}}
\put(   63.50000,2){\line(0,1){    .00000}}


\put(-0.8,0.5){$\scriptscriptstyle{-\frac{10}{100}}$}
\put(16.2,0.5){$\scriptscriptstyle{\frac{10}{100}}$}
\put(22.2,0.5){$\scriptscriptstyle{-\frac{10}{100}}$}
\put(39.2,0.5){$\scriptscriptstyle{\frac{10}{100}}$}
\put(45.2,0.5){$\scriptscriptstyle{-\frac{10}{100}}$}
\put(62.2,0.5){$\scriptscriptstyle{\frac{10}{100}}$}
\put(3.5,0.5){$\scriptscriptstyle{-\frac{5}{100}}$}
\put(12.3,0.5){$\scriptscriptstyle{\frac{5}{100}}$}
\put(26.5,0.5){$\scriptscriptstyle{-\frac{5}{100}}$}
\put(35.3,0.5){$\scriptscriptstyle{\frac{5}{100}}$}
\put(49.5,0.5){$\scriptscriptstyle{-\frac{5}{100}}$}
\put(58.3,0.5){$\scriptscriptstyle{\frac{5}{100}}$}

\put(6.7,13.5){$\scriptscriptstyle{0.75}$}
\put( 9.3,14){\line(1,0){0.4}}
\put(7.2,9.5){$\scriptscriptstyle{0.5}$}
\put( 9.3,10){\line(1,0){0.4}}
\put(19,2.5){$\scriptstyle{x}$}
\put(10,16){$\scriptstyle{\gamma _{0.2}(x)}$}

\put(29.7,13.5){$\scriptscriptstyle{0.75}$}
\put( 32.3,14){\line(1,0){0.4}}
\put(42,2.5){$\scriptstyle{x}$}
\put(33,16){$\scriptstyle{\gamma _{1}(x)}$}

\put(52.7,13.5){$\scriptscriptstyle{0.75}$}
\put( 55.3,14){\line(1,0){0.4}}
\put(65,2.5){$\scriptstyle{x}$}
\put(56,16){$\scriptstyle{\gamma _{2}(x)}$}
\end{picture}
\caption{The functions $\gamma _{0.2}$ (left),  $\gamma _1$ (center) and $\gamma _2$ (right) in the case $q=10$.}\label{gaussians}
\end{figure}
\end{center}

\section{Market information and the price evolution}
\label{evolution}

The rate of return of the stock market in equilibrium is usually described by a Gaussian function \cite{Zhang,Cootner}. In the case of our model we use a function 
 \begin{equation}
\gamma _\alpha   :\mathcal{L}_d\longrightarrow \mathbb{R},\qquad \gamma _\alpha  (x)=\frac{1}{\sqrt{\langle g_\alpha ,g_\alpha \rangle }}\, g_\alpha (x)
\end{equation}
(see Figure \ref{gaussians}) depending on a parameter $\alpha \!\in \!(0,\infty )$, where
\begin{equation}
\begin{array}{l}
g_\alpha \left(\frac{n}{100}\right)=\sum_{m =-\infty }^\infty \mathrm{e}^{-\frac{\alpha  \pi }{d} (m d +n)^2}=\frac{1}{\sqrt{\alpha d}}\, \theta _3\left( \frac{n}{d},\frac{{\rm i}}{\alpha d} \right).
\end{array}
\end{equation}
By using the properties of the Jacobi function
\begin{equation}
\theta _3(z,\tau )=\sum_{\alpha =-\infty }^\infty {\rm e}^{{\rm i}\pi \tau \alpha ^2}\, {\rm e}^{2\pi {\rm i}\alpha z}
\end{equation}
Ruzzi \cite{Ruzzi} has obtained the relation
\begin{equation}
F[g _\alpha ]=\frac{1}{\sqrt{\alpha }}\, g _{\frac{1}{\alpha }}.
\end{equation}

In the case of a Hamiltonian
\begin{equation}
\begin{array}{l}
\hat H=\frac{1}{2\mu }\hat T^2+\mathcal{V}(\hat R,t)
\end{array}
\end{equation}
the kinetic part $\frac{1}{2\mu }\hat T^2$ represents the efforts of the traders to change prices \cite{Choustova}. The intensive exchange of information in the world of finances is one of the main sources determining dynamics of prices. The potential part $\mathcal{V}(\hat R,t)$ of $\hat H$ describes the interactions between traders as well as external economic conditions and even meteorological conditions \cite{Choustova}.

In the particular case 
\[
\begin{array}{l}
\hat H=\frac{1}{2\mu }\hat T^2
\end{array}
\]
the solution satisfying the initial condition $\Phi \left(\frac{n}{100},0\right)=\varphi \left(\frac{n}{100}\right)$ is
\[  
\begin{array}{l}
\Phi \left(\frac{n}{100},t\right) ={\rm e}^{-i\frac{t}{2\mu }\hat T^2}\varphi \left(\frac{n}{100}\right).
\end{array}
\]
From the relation
\[
\begin{array}{rl}
{\rm e}^{-i\frac{t}{2\mu }\hat T^2}\varphi  & =\sum\limits_{k=-q}^qF[\varphi ]\left(\frac{k}{100}\right) \,{\rm e}^{-i\frac{t}{2\mu }\hat T^2}\Phi _k\\[3mm]
 & =\sum\limits_{k=-q}^qF[\varphi ]\left(\frac{k}{100}\right) \,{\rm e}^{-i\frac{t}{2\mu }\frac{k^2}{10000}}\Phi _k
\end{array}
\]
obtained from (\ref{Teigen})  and (\ref{Trepr}), it follows that
\[
\begin{array}{l}
{\rm e}^{-i\frac{t}{2\mu }\hat T^2}\varphi  ={\rm e}^{-i\frac{t+40000\pi \mu }{2\mu }\hat T^2}\varphi 
\end{array}
\]
whence
\[  
\begin{array}{l}
\Phi \left(\frac{n}{100},t\right) =\left\langle \delta _n,{\rm e}^{-i\frac{t}{2\mu }\hat T^2}\varphi  \right\rangle =\Phi \left(\frac{n}{100},t+40000\pi \mu \right)
\end{array}
\]
that is, the time evolution is periodic with a period of $40 000\pi \mu $ seconds.
This mathematical result, which does not correspond to a realistic case, shows that  $\mathcal{V}(\hat R,t)$ plays a key role in the description of the price evolution.

The total effect of market information affecting the stock price at a certain time determines either the stock price's rise or the stock price's decline. In order to illustrate our model, by following \cite{Zhang}, we consider an idealized model in which we assume two type of information appear periodically. We choose a cosine function $\cos \omega t$ to simulate the fluctuation of the information and use the Hamiltonian
\begin{equation}
\begin{array}{l}
\hat H=\frac{1}{2\mu }\hat T^2+\beta \, \hat R \cos \omega t.
\end{array}
\end{equation} 
where $\beta $ is a constant.
The solution of the Schr\" odinger equation 
\begin{equation}
{\rm i}\, \frac{\partial \Phi }{\partial t}=\left[ \frac{1}{2\mu }\hat T^2+\beta \, \hat R \cos \omega t\right] \Phi 
\end{equation}
can be obtained by using a program in Mathematica (see the Appendix).
We assume that at the opening time ($t=0$) of the stock market, the wave function describing the rate of return is the function $\gamma _\alpha $ (corresponding to a certain $\alpha $), that is,
\begin{equation}
\begin{array}{l}
\Phi \left(\frac{n}{100},0\right)=\gamma _\alpha \left(\frac{n}{100}\right).
\end{array}
\end{equation}
The role played by the kinetic part of the Hamiltonian and the role played by the market information depend on the values of the parameters $\mu $, $\beta $ and $\omega $.
The distribution of the probabilities $|\Psi (n,t)|^2$ corresponding to the possible values $\frac{n}{100}$ of the rate of return at certain moments of time are presented in the figure 2.
We start from a stock market in equilibrium with the rate of return described by the Gaussian $\gamma _{0.2}$. After 1800 seconds the most probable rate of return is $-2\%$ and after 3600 seconds  $-5\%$ and $-4\%$. The rate of return comes back to equilibrium after 7200 seconds. After 14400 seconds the values $-3\%$ and $-4\%$ are more probable than any other value. After 28800 seconds the most probable rate of return is $4\%$.

\begin{center}
\begin{figure}[t]
\setlength{\unitlength}{1.8mm}
\begin{picture}(70,40)(-5,0)
\put(0,2){\vector(1,0){20}}
\put(23,2){\vector(1,0){20}}
\put(46,2){\vector(1,0){20}}
\put(9.5,0){\vector(0,1){17}}
\put(32.5,0){\vector(0,1){17}}
\put(55.5,0){\vector(0,1){17}}

\put(0,22){\vector(1,0){20}}
\put(23,22){\vector(1,0){20}}
\put(46,22){\vector(1,0){20}}
\put(9.5,20){\vector(0,1){17}}
\put(32.5,20){\vector(0,1){17}}
\put(55.5,20){\vector(0,1){17}}

\put(    1.50000,  22.03669){\circle*{0.3}}
\put(    2.30000,  22.06452){\circle*{0.3}}
\put(    3.10000,  22.14637){\circle*{0.3}}
\put(    3.90000,  22.33818){\circle*{0.3}}
\put(    4.70000,  22.72346){\circle*{0.3}}
\put(    5.50000,  23.39031){\circle*{0.3}}
\put(    6.30000,  24.37895){\circle*{0.3}}
\put(    7.10000,  25.61516){\circle*{0.3}}
\put(    7.90000,  26.87548){\circle*{0.3}}
\put(    8.70000,  27.83402){\circle*{0.3}}
\put(    9.50000,  28.19375){\circle*{0.3}}
\put(   10.30000,  27.83402){\circle*{0.3}}
\put(   11.10000,  26.87548){\circle*{0.3}}
\put(   11.90000,  25.61516){\circle*{0.3}}
\put(   12.70000,  24.37895){\circle*{0.3}}
\put(   13.50000,  23.39031){\circle*{0.3}}
\put(   14.30000,  22.72346){\circle*{0.3}}
\put(   15.10000,  22.33818){\circle*{0.3}}
\put(   15.90000,  22.14637){\circle*{0.3}}
\put(   16.70000,  22.06452){\circle*{0.3}}
\put(   17.50000,  22.03669){\circle*{0.3}}
\put(    1.50000,22){\line(0,1){    .03669}}
\put(    2.30000,22){\line(0,1){    .06452}}
\put(    3.10000,22){\line(0,1){    .14637}}
\put(    3.90000,22){\line(0,1){    .33818}}
\put(    4.70000,22){\line(0,1){    .72346}}
\put(    5.50000,22){\line(0,1){   1.39031}}
\put(    6.30000,22){\line(0,1){   2.37895}}
\put(    7.10000,22){\line(0,1){   3.61516}}
\put(    7.90000,22){\line(0,1){   4.87548}}
\put(    8.70000,22){\line(0,1){   5.83403}}
\put(    9.50000,22){\line(0,1){   6.19376}}
\put(   10.30000,22){\line(0,1){   5.83403}}
\put(   11.10000,22){\line(0,1){   4.87548}}
\put(   11.90000,22){\line(0,1){   3.61516}}
\put(   12.70000,22){\line(0,1){   2.37895}}
\put(   13.50000,22){\line(0,1){   1.39031}}
\put(   14.30000,22){\line(0,1){    .72346}}
\put(   15.10000,22){\line(0,1){    .33818}}
\put(   15.90000,22){\line(0,1){    .14637}}
\put(   16.70000,22){\line(0,1){    .06452}}
\put(   17.50000,22){\line(0,1){    .03669}}
\put(   24.50000,  22.23069){\circle*{0.3}}
\put(   25.30000,  22.31801){\circle*{0.3}}
\put(   26.10000,  22.59318){\circle*{0.3}}
\put(   26.90000,  23.29181){\circle*{0.3}}
\put(   27.70000,  24.16144){\circle*{0.3}}
\put(   28.50000,  25.37613){\circle*{0.3}}
\put(   29.30000,  26.62276){\circle*{0.3}}
\put(   30.10000,  27.65582){\circle*{0.3}}
\put(   30.90000,  28.13359){\circle*{0.3}}
\put(   31.70000,  27.91867){\circle*{0.3}}
\put(   32.50000,  27.07290){\circle*{0.3}}
\put(   33.30000,  25.87041){\circle*{0.3}}
\put(   34.10000,  24.62170){\circle*{0.3}}
\put(   34.90000,  23.57189){\circle*{0.3}}
\put(   35.70000,  22.83793){\circle*{0.3}}
\put(   36.50000,  22.41562){\circle*{0.3}}
\put(   37.30000,  22.20245){\circle*{0.3}}
\put(   38.10000,  22.08433){\circle*{0.3}}
\put(   38.90000,  22.01885){\circle*{0.3}}
\put(   39.70000,  22.00133){\circle*{0.3}}
\put(   40.50000,  22.00047){\circle*{0.3}}
\put(   24.50000,22){\line(0,1){    .23069}}
\put(   25.30000,22){\line(0,1){    .31801}}
\put(   26.10000,22){\line(0,1){    .59318}}
\put(   26.90000,22){\line(0,1){   1.29181}}
\put(   27.70000,22){\line(0,1){   2.16144}}
\put(   28.50000,22){\line(0,1){   3.37613}}
\put(   29.30000,22){\line(0,1){   4.62276}}
\put(   30.10000,22){\line(0,1){   5.65583}}
\put(   30.90000,22){\line(0,1){   6.13359}}
\put(   31.70000,22){\line(0,1){   5.91867}}
\put(   32.50000,22){\line(0,1){   5.07290}}
\put(   33.30000,22){\line(0,1){   3.87041}}
\put(   34.10000,22){\line(0,1){   2.62170}}
\put(   34.90000,22){\line(0,1){   1.57189}}
\put(   35.70000,22){\line(0,1){    .83793}}
\put(   36.50000,22){\line(0,1){    .41563}}
\put(   37.30000,22){\line(0,1){    .20245}}
\put(   38.10000,22){\line(0,1){    .08433}}
\put(   38.90000,22){\line(0,1){    .01885}}
\put(   39.70000,22){\line(0,1){    .00133}}
\put(   40.50000,22){\line(0,1){    .00047}}
\put(   47.50000,  22.06784){\circle*{0.3}}
\put(   48.30000,  23.32140){\circle*{0.3}}
\put(   49.10000,  25.78778){\circle*{0.3}}
\put(   49.90000,  26.75020){\circle*{0.3}}
\put(   50.70000,  27.32003){\circle*{0.3}}
\put(   51.50000,  28.03275){\circle*{0.3}}
\put(   52.30000,  28.10038){\circle*{0.3}}
\put(   53.10000,  27.56767){\circle*{0.3}}
\put(   53.90000,  26.49599){\circle*{0.3}}
\put(   54.70000,  25.23861){\circle*{0.3}}
\put(   55.50000,  24.06644){\circle*{0.3}}
\put(   56.30000,  23.17644){\circle*{0.3}}
\put(   57.10000,  22.59823){\circle*{0.3}}
\put(   57.90000,  22.27615){\circle*{0.3}}
\put(   58.70000,  22.11742){\circle*{0.3}}
\put(   59.50000,  22.04747){\circle*{0.3}}
\put(   60.30000,  22.01923){\circle*{0.3}}
\put(   61.10000,  22.00894){\circle*{0.3}}
\put(   61.90000,  22.00434){\circle*{0.3}}
\put(   62.70000,  22.00083){\circle*{0.3}}
\put(   63.50000,  22.00188){\circle*{0.3}}
\put(   47.50000,22){\line(0,1){    .06784}}
\put(   48.30000,22){\line(0,1){   1.32140}}
\put(   49.10000,22){\line(0,1){   3.78778}}
\put(   49.90000,22){\line(0,1){   4.75020}}
\put(   50.70000,22){\line(0,1){   5.32003}}
\put(   51.50000,22){\line(0,1){   6.03274}}
\put(   52.30000,22){\line(0,1){   6.10038}}
\put(   53.10000,22){\line(0,1){   5.56767}}
\put(   53.90000,22){\line(0,1){   4.49599}}
\put(   54.70000,22){\line(0,1){   3.23861}}
\put(   55.50000,22){\line(0,1){   2.06644}}
\put(   56.30000,22){\line(0,1){   1.17644}}
\put(   57.10000,22){\line(0,1){    .59823}}
\put(   57.90000,22){\line(0,1){    .27615}}
\put(   58.70000,22){\line(0,1){    .11742}}
\put(   59.50000,22){\line(0,1){    .04747}}
\put(   60.30000,22){\line(0,1){    .01923}}
\put(   61.10000,22){\line(0,1){    .00894}}
\put(   61.90000,22){\line(0,1){    .00434}}
\put(   62.70000,22){\line(0,1){    .00083}}
\put(   63.50000,22){\line(0,1){    .00188}}
\put(    1.50000,   2.32819){\circle*{0.3}}
\put(    2.30000,   2.03512){\circle*{0.3}}
\put(    3.10000,   2.16030){\circle*{0.3}}
\put(    3.90000,   2.31371){\circle*{0.3}}
\put(    4.70000,   2.55017){\circle*{0.3}}
\put(    5.50000,   3.03455){\circle*{0.3}}
\put(    6.30000,   3.86832){\circle*{0.3}}
\put(    7.10000,   5.04204){\circle*{0.3}}
\put(    7.90000,   6.35401){\circle*{0.3}}
\put(    8.70000,   7.49450){\circle*{0.3}}
\put(    9.50000,   8.11429){\circle*{0.3}}
\put(   10.30000,   8.04431){\circle*{0.3}}
\put(   11.10000,   7.26329){\circle*{0.3}}
\put(   11.90000,   6.12071){\circle*{0.3}}
\put(   12.70000,   4.78765){\circle*{0.3}}
\put(   13.50000,   3.75958){\circle*{0.3}}
\put(   14.30000,   2.92920){\circle*{0.3}}
\put(   15.10000,   2.42702){\circle*{0.3}}
\put(   15.90000,   2.26080){\circle*{0.3}}
\put(   16.70000,   2.10534){\circle*{0.3}}
\put(   17.50000,   2.00691){\circle*{0.3}}
\put(    1.50000,2){\line(0,1){    .32819}}
\put(    2.30000,2){\line(0,1){    .03512}}
\put(    3.10000,2){\line(0,1){    .16029}}
\put(    3.90000,2){\line(0,1){    .31371}}
\put(    4.70000,2){\line(0,1){    .55017}}
\put(    5.50000,2){\line(0,1){   1.03455}}
\put(    6.30000,2){\line(0,1){   1.86832}}
\put(    7.10000,2){\line(0,1){   3.04204}}
\put(    7.90000,2){\line(0,1){   4.35401}}
\put(    8.70000,2){\line(0,1){   5.49450}}
\put(    9.50000,2){\line(0,1){   6.11428}}
\put(   10.30000,2){\line(0,1){   6.04431}}
\put(   11.10000,2){\line(0,1){   5.26329}}
\put(   11.90000,2){\line(0,1){   4.12071}}
\put(   12.70000,2){\line(0,1){   2.78765}}
\put(   13.50000,2){\line(0,1){   1.75958}}
\put(   14.30000,2){\line(0,1){    .92920}}
\put(   15.10000,2){\line(0,1){    .42702}}
\put(   15.90000,2){\line(0,1){    .26080}}
\put(   16.70000,2){\line(0,1){    .10534}}
\put(   17.50000,2){\line(0,1){    .00691}}
\put(   24.50000,   2.22142){\circle*{0.3}}
\put(   25.30000,   2.15772){\circle*{0.3}}
\put(   26.10000,   2.03522){\circle*{0.3}}
\put(   26.90000,   2.10466){\circle*{0.3}}
\put(   27.70000,   3.14867){\circle*{0.3}}
\put(   28.50000,   3.17518){\circle*{0.3}}
\put(   29.30000,   6.15439){\circle*{0.3}}
\put(   30.10000,  13.43886){\circle*{0.3}}
\put(   30.90000,  13.10236){\circle*{0.3}}
\put(   31.70000,   8.50992){\circle*{0.3}}
\put(   32.50000,   5.60920){\circle*{0.3}}
\put(   33.30000,   4.22168){\circle*{0.3}}
\put(   34.10000,   3.37579){\circle*{0.3}}
\put(   34.90000,   2.74373){\circle*{0.3}}
\put(   35.70000,   2.35260){\circle*{0.3}}
\put(   36.50000,   2.22090){\circle*{0.3}}
\put(   37.30000,   2.12321){\circle*{0.3}}
\put(   38.10000,   2.03390){\circle*{0.3}}
\put(   38.90000,   2.08147){\circle*{0.3}}
\put(   39.70000,   2.05860){\circle*{0.3}}
\put(   40.50000,   2.13052){\circle*{0.3}}
\put(   24.50000,2){\line(0,1){    .22142}}
\put(   25.30000,2){\line(0,1){    .15772}}
\put(   26.10000,2){\line(0,1){    .03522}}
\put(   26.90000,2){\line(0,1){    .10466}}
\put(   27.70000,2){\line(0,1){   1.14867}}
\put(   28.50000,2){\line(0,1){   1.17518}}
\put(   29.30000,2){\line(0,1){   4.15439}}
\put(   30.10000,2){\line(0,1){  11.43886}}
\put(   30.90000,2){\line(0,1){  11.10236}}
\put(   31.70000,2){\line(0,1){   6.50992}}
\put(   32.50000,2){\line(0,1){   3.60920}}
\put(   33.30000,2){\line(0,1){   2.22168}}
\put(   34.10000,2){\line(0,1){   1.37579}}
\put(   34.90000,2){\line(0,1){    .74373}}
\put(   35.70000,2){\line(0,1){    .35260}}
\put(   36.50000,2){\line(0,1){    .22090}}
\put(   37.30000,2){\line(0,1){    .12321}}
\put(   38.10000,2){\line(0,1){    .03390}}
\put(   38.90000,2){\line(0,1){    .08147}}
\put(   39.70000,2){\line(0,1){    .05860}}
\put(   40.50000,2){\line(0,1){    .13052}}
\put(   47.50000,   2.01781){\circle*{0.3}}
\put(   48.30000,   2.17846){\circle*{0.3}}
\put(   49.10000,   2.00218){\circle*{0.3}}
\put(   49.90000,   2.23587){\circle*{0.3}}
\put(   50.70000,   2.50557){\circle*{0.3}}
\put(   51.50000,   3.03165){\circle*{0.3}}
\put(   52.30000,   2.67882){\circle*{0.3}}
\put(   53.10000,   2.12255){\circle*{0.3}}
\put(   53.90000,   2.31818){\circle*{0.3}}
\put(   54.70000,   3.21771){\circle*{0.3}}
\put(   55.50000,   2.48644){\circle*{0.3}}
\put(   56.30000,   4.70926){\circle*{0.3}}
\put(   57.10000,   5.97091){\circle*{0.3}}
\put(   57.90000,   8.96429){\circle*{0.3}}
\put(   58.70000,   9.95978){\circle*{0.3}}
\put(   59.50000,   8.08197){\circle*{0.3}}
\put(   60.30000,   7.51983){\circle*{0.3}}
\put(   61.10000,   6.60327){\circle*{0.3}}
\put(   61.90000,   3.63150){\circle*{0.3}}
\put(   62.70000,   2.60217){\circle*{0.3}}
\put(   63.50000,   2.16176){\circle*{0.3}}
\put(   47.50000,2){\line(0,1){    .01781}}
\put(   48.30000,2){\line(0,1){    .17846}}
\put(   49.10000,2){\line(0,1){    .00218}}
\put(   49.90000,2){\line(0,1){    .23587}}
\put(   50.70000,2){\line(0,1){    .50558}}
\put(   51.50000,2){\line(0,1){   1.03165}}
\put(   52.30000,2){\line(0,1){    .67882}}
\put(   53.10000,2){\line(0,1){    .12255}}
\put(   53.90000,2){\line(0,1){    .31818}}
\put(   54.70000,2){\line(0,1){   1.21771}}
\put(   55.50000,2){\line(0,1){    .48644}}
\put(   56.30000,2){\line(0,1){   2.70926}}
\put(   57.10000,2){\line(0,1){   3.97091}}
\put(   57.90000,2){\line(0,1){   6.96429}}
\put(   58.70000,2){\line(0,1){   7.95978}}
\put(   59.50000,2){\line(0,1){   6.08198}}
\put(   60.30000,2){\line(0,1){   5.51983}}
\put(   61.10000,2){\line(0,1){   4.60327}}
\put(   61.90000,2){\line(0,1){   1.63150}}
\put(   62.70000,2){\line(0,1){    .60217}}
\put(   63.50000,2){\line(0,1){    .16176}}

\put(-0.8,20.5){$\scriptscriptstyle{-\frac{10}{100}}$}
\put(16.2,20.5){$\scriptscriptstyle{\frac{10}{100}}$}
\put(22.2,20.5){$\scriptscriptstyle{-\frac{10}{100}}$}
\put(39.2,20.5){$\scriptscriptstyle{\frac{10}{100}}$}
\put(45.2,20.5){$\scriptscriptstyle{-\frac{10}{100}}$}
\put(62.2,20.5){$\scriptscriptstyle{\frac{10}{100}}$}

\put(2,35){$\scriptscriptstyle{t=0\, s}$}
\put(10,33){$\scriptscriptstyle{0.25}$}
\put( 9.3,33.2){\line(1,0){0.4}}
\put(10,28.5){$\scriptscriptstyle{0.15}$}
\put( 9.3,28.75){\line(1,0){0.4}}
\put(19,22.5){$\scriptstyle{x}$}
\put(10,36){$\scriptstyle{|\Phi(x,t)|^2}$}

\put(25,35){$\scriptscriptstyle{t=1800\, s}$}
\put(33,33){$\scriptscriptstyle{0.25}$}
\put( 32.3,33.2){\line(1,0){0.4}}
\put(33,28.3){$\scriptscriptstyle{0.15}$}
\put( 32.3,28.75){\line(1,0){0.4}}
\put(42,22.5){$\scriptstyle{x}$}
\put(33,36){$\scriptstyle{|\Phi(x,t)|^2}$}

\put(48,35){$\scriptscriptstyle{t=3600\, s}$}
\put(56,33){$\scriptscriptstyle{0.25}$}
\put( 55.3,33.2){\line(1,0){0.4}}
\put(56,28.3){$\scriptscriptstyle{0.15}$}
\put( 55.3,28.75){\line(1,0){0.4}}
\put(65,22.5){$\scriptstyle{x}$}
\put(56,36){$\scriptstyle{|\Phi(x,t)|^2}$}
\put(-0.8,0.5){$\scriptscriptstyle{-\frac{10}{100}}$}
\put(16.2,0.5){$\scriptscriptstyle{\frac{10}{100}}$}
\put(22.2,0.5){$\scriptscriptstyle{-\frac{10}{100}}$}
\put(40,0.5){$\scriptscriptstyle{\frac{10}{100}}$}
\put(45.2,0.5){$\scriptscriptstyle{-\frac{10}{100}}$}
\put(63,0.5){$\scriptscriptstyle{\frac{10}{100}}$}
\put(3.5,0.5){$\scriptscriptstyle{-\frac{5}{100}}$}
\put(12.3,0.5){$\scriptscriptstyle{\frac{5}{100}}$}
\put(26.5,0.5){$\scriptscriptstyle{-\frac{5}{100}}$}
\put(35.3,0.5){$\scriptscriptstyle{\frac{5}{100}}$}
\put(49.5,0.5){$\scriptscriptstyle{-\frac{5}{100}}$}
\put(58.3,0.5){$\scriptscriptstyle{\frac{5}{100}}$}

\put(3.5,20.5){$\scriptscriptstyle{-\frac{5}{100}}$}
\put(12.3,20.5){$\scriptscriptstyle{\frac{5}{100}}$}
\put(26.5,20.5){$\scriptscriptstyle{-\frac{5}{100}}$}
\put(35.3,20.5){$\scriptscriptstyle{\frac{5}{100}}$}
\put(49.5,20.5){$\scriptscriptstyle{-\frac{5}{100}}$}
\put(58.3,20.5){$\scriptscriptstyle{\frac{5}{100}}$}

\put(2,15){$\scriptscriptstyle{t=7200\, s}$}
\put(10,13){$\scriptscriptstyle{0.25}$}
\put( 9.3,13.2){\line(1,0){0.4}}
\put(10,8.7){$\scriptscriptstyle{0.15}$}
\put( 9.3,8.75){\line(1,0){0.4}}
\put(19,2.5){$\scriptstyle{x}$}
\put(10,16){$\scriptstyle{|\Phi(x,t)|^2}$}

\put(25,15){$\scriptscriptstyle{t=14400\, s}$}
\put(33,13){$\scriptscriptstyle{0.25}$}
\put( 32.3,13.2){\line(1,0){0.4}}
\put(33,8.3){$\scriptscriptstyle{0.15}$}
\put( 32.3,8.75){\line(1,0){0.4}}
\put(42,2.5){$\scriptstyle{x}$}
\put(33,16){$\scriptstyle{|\Phi(x,t)|^2}$}

\put(48,15){$\scriptscriptstyle{t=28800\, s}$}
\put(56,13){$\scriptscriptstyle{0.25}$}
\put( 55.3,13.2){\line(1,0){0.4}}
\put(52.5,8.3){$\scriptscriptstyle{0.15}$}
\put( 55.3,8.75){\line(1,0){0.4}}
\put(65,2.5){$\scriptstyle{x}$}
\put(56,16){$\scriptstyle{|\Phi(x,t)|^2}$}
\end{picture}
\caption{The probability $|\Phi \left(\frac{n}{100},t\right)|^2$ to have a rate of return equal to $n\%$ for $t\!=\!0$, 1800, 3600, 7200, 14400 and  28800 in the particular case $\alpha \!=\!0.2$, $\mu \!=\!1$, $\beta \!=\!1/10$, $\omega \!=\!1/10000$.}
\end{figure}
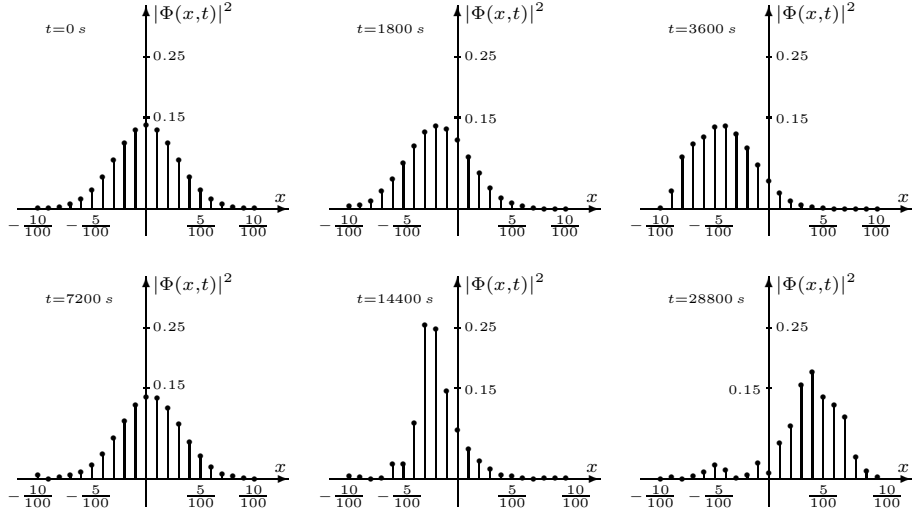
\end{center}

\section{Conclusion}
\label{conclusion}

Our finite-dimensional version,  based on the mathematical formalism used in the case of finite quantum systems, keeps the essential characteristics of the model proposed by Zhang and Huang in \cite{Zhang}. It takes into consideration the discrete nature of the quantities used in the field of finance, and the mathematical formalism is simpler:
\begin{itemize}
\item[-] we use finite sums instead of integrals

\item[-] in our approach any non-null function can be normalized

\item[-] the spectra of our operators are finite sets

\item[-] our operators are defined on the whole Hilbert space

\item[-] the notion of Hermitian operator can be defined in a simpler way 

\item[-] the Schr\" odinger type equation we use is equivalent to a system of $d$ first order differential linear equations with $d$ unknown functions. The Schr\" odinger equation used in \cite{Zhang} is a second order differential equation with partial derivatives.
\end{itemize}
\bigskip
\appendix{\bf Appendix}\\
\label{appendix}

\noindent The time evolution of the probabilities $|\Phi (n,t)|^2$ corresponding to the possible values $\frac{n}{100}$ of the rate of return can be obtained by using the following program in Mathematica:
{\scriptsize

\begin{verbatim}
time = 3600;   alpha = 0.2; mu = 1; beta = 1/10; omega =  1/10000; d = 21; q := (d - 1)/2 
g[n_] := N[  Sum[Exp[-Pi  alpha (m d + n)^2/d], {m, -Infinity, Infinity}]]
gamma := Normalize[Table[ g[n], {n, -q, q}]]
T[n_, m_] := N[(1/(100 d) ) Sum[k Exp[2 Pi I (m - n) k/d], {k, -q, q}]]
H[n_, m_, t_] := (1/(2 mu)) Sum[ T[n, k] T[k, m], {k, -q, q}]  
                                         + beta (n/100) Cos[omega t] DiscreteDelta[n - m]
eqns = {Table[   phi[n]'[t] == Sum[- I H[n, m, t] phi[m][t], {m, 1, d}], {n, 1, d}], 
                                          Table[phi[n][0] == gamma[[n]], {n, 1, d}]}
ndsolve = NDSolve[eqns, Table[phi[n], {n, d}], {t, 30000}]
Phi = Normalize[  Table[Evaluate[Abs[phi[n][time]] /. ndsolve][[1]], {n, 1, d}]]
ListPlot[Table[{n, Phi[[n + q + 1]]^2}, {n, -q, q}] , Filling -> Axis]
\end{verbatim}
}
\noindent The reader can easily change the values of parameters \verb#time#, \verb#alpha #,  \verb#mu#,  \verb#beta# and the function \verb#Cos[omega t]# in order to investigate other cases.








\end{document}